\documentclass[numberedappendix,apjl]{emulateapj}
\usepackage{mathptmx}

\begin{document}

\title{The Dust Content of Galaxy Clusters}
\author{Doron Chelouche\altaffilmark{1,2}, Benjamin P. Koester\altaffilmark{3}, and David V. Bowen  \altaffilmark{4}}
\altaffiltext{1} {School of Natural Sciences, Institute for
Advanced Study, Einstein Drive, Princeton 08540, USA;
doron@ias.edu} \altaffiltext{2} {Chandra Fellow} \altaffiltext{3} {Kavli Institute for Cosmological Physics, Department of Astronomy and Astrophysics, The University of Chicago, 5640 South Ellis Avenue, Chicago, IL 60637; bkoester@oddjob.uchicago.edu}  \altaffiltext{4} {Department of Astrophysical Sciences, Princeton University, Princeton, NJ 08544; dvb@astro.princeton.edu}

\shortauthors{Chelouche D. et al.}
\shorttitle{Dust in  Galaxy Clusters}

\begin{abstract}

We report on the detection of reddening toward $z\sim 0.2$ galaxy clusters. This is measured by correlating the {\it Sloan Digital Sky Survey} cluster and quasar catalogs and by comparing the photometric and spectroscopic properties of quasars behind the clusters to those in the field. We find mean $E(B-V)$ values of a few$\times 10^{-3}$\,mag for sight lines passing $\sim$Mpc from the clusters' center. The reddening curve is typical of dust but cannot be used to distinguish between different dust types. The radial dependence of the extinction is shallow near the cluster center suggesting that most of the detected dust lies at the outskirts of the clusters.  Gravitational magnification of background $z\sim 1.7$ sources seen on Mpc (projected) scales around the clusters is found to be of order a few per cent, in qualitative agreement with theoretical predictions.  Contamination by different spectral properties of the lensed quasar population is unlikely but cannot be excluded.
\end{abstract}


\section{Introduction}

 Zwicky (1957) suggested the presence of dust  in the intracluster medium (ICM) to explain galaxy number counts behind the Coma cluster and a search has begun to quantify its properties. Similar analyses were carried out by Karachentsev \& Lipovetskii (1969), Bogart \& Wagoner (1973), and Boyle et al. (1988 using quasars)  and yielded B-band extinctions of order 0.2\,mag.  However, a more recent study by Maoz (1993) using radio-selected quasars behind rich Abell clusters yielded only upper-limits on the reddening [$E(B-V)<0.05$\,mag] with similar limits [$E(B-V)< 0.02$\,mag] obtained from the analysis of galaxy colors behind APM clusters (e.g., Nollenberg et al. 2003 see also Ferguson 1993). In contrast, Hu (1992) found $E(B-V)\sim 0.2$\,mag toward cooling flow clusters (but see Annis \& Jewitt 1993).  Dust emission from clusters has been marginally detected in the infrared  resulting in a dust-to-gas ratio as low as $\sim 10^{-6}$ in a few Abell clusters (Stickel et al.  2002; but see Bai et al. 2007).
 
Cluster members are known to contain dust (e.g., Edge et al. 1999, Popescu et al.  2002, Boselli et al. 2004) and galaxies are seen to shed gas and dust into their environments (e.g., Heckman et al. 2000; see also Jonsson et al. 2006). However, the survival time for small dust grains mixed with the ICM is much shorter than Hubble time (Draine \& Salpeter 1979), and would be absent from the cluster core (but could survive on large scales; see Dwek et al. 1990) unless constantly replenished (e.g., Montier \& Giard 2004) or  shielded from the hot plasma (e.g., Voit \& Donahue 1995).   Presently, there are few meaningful constraints on the composition and quantity of intracluster dust, an understanding of which would have important implications for cluster evolution and metal enrichment processes, as well as for dust physics. In this {\it Letter} we provide new answers to 50 years old questions concerning dust in galaxy clusters, by harnessing the power of the {\it Sloan Digital Sky Survey} (SDSS).

\section{Method}

We use a $0.1<z<0.3$ sample of $\sim 10^4$ optically-selected clusters from
Koester et al. (2007) with velocity dispersions $>400~{\rm
km~s^{-1}}$.  The mean $R_{200}$ (the radius within which the galaxy density is $200\times$ the field density) is $\sim$1\,Mpc\footnote{We use the standard cosmology with $H_0=70~{\rm km~s^{-1} Mpc^{-1}},~\Omega_m=0.3$, and $\Omega_\Lambda=0.7$}. The cluster center is defined by its brightest member whose position may deviate from the center of mass by a few$\times 100$\,kpc.   This sample was correlated with the SDSS/DR5 spectroscopic quasar catalog (Schneider et al. 2007), identifying all quasars within 10\,Mpc of the clusters' centers. All data were corrected for galactic extinction using the  values of Schneider et al. (2007). In particular, the spectra were corrected using the reddening curve of Cardelli et al. (1989). 

We employ a statistical approach to search for dust by comparing the photometric and spectroscopic properties of quasars at different impact parameters, $b$, from the cluster center with those of quasars in the field. The SDSS quasar selection criteria have no effect on the results and do not bias the observed colors for $E(B-V)\sim 0.7E(g-i)<0.1$\,mag (see M\'enard et al. 2007). As clusters vary in size, we define a normalized impact parameter:
\begin{equation}
\tilde{b} \equiv {b}/{R_{200}}.
\end{equation}
A cluster  would induce a mean color excess in background objects whose sight lines pass  a distance $\tilde{b}$ from its center (see appendix A):
\begin{equation}
 \left < E(i-j) \right >_{\tilde{b}}  =  \frac{2.5} {{\rm ln} 10} \left < \tau_i-\tau_j \right >_{\tilde{b}} + \left < \delta^F_i-\delta^F_j \right >_{\tilde{b}} + \xi_{\tilde{b}}(\mu), 
\end{equation}
where $i,j$ denote SDSS photometric bands, $\tau$  the effective dust optical depth, and $\delta^F$  the foreground emission by the cluster. These terms result in effective reddening each having very different signatures (see Fig. 1 where median quasar/early-type cluster galaxy colors were used so that $[u,g,r,i,z]_{\rm quasar/galaxy}=[19.4,19.2,19.1,19.0,18.9]/[21.2,19.9,18.8,18.3,18.0]$).  The $\xi(\mu)$ term arises when quasar colors are luminosity dependent and gravitational magnification, $\mu$ is considerable (see \S 3 and the appendix). 

The analysis proceeds as follows: we define an annulus about the
cluster center in the range $[\tilde{b}, \tilde{b}+\delta \tilde{b}]$. All quasars within this
annulus serve as a primary sample (consisting of  $\sim 3000$ objects for $\tilde{b} \in [0, 1]$ with the number $\propto \tilde{b}\delta \tilde{b}$). Many primary samples span the range $\tilde{b} \in [0,7]$ with $\delta \tilde{b}$ being set by the
requirement for adequate S/N and spatial resolution.  We define our control sample to be all quasars with $\tilde{b}>7$. 
As the control sample is large, we choose many random control sub-samples with the same number of objects as the primary sample and compare the geometric mean of their spectral and photometric properties. This allows us to quantify the mean color excess and its variance  via Monte Carlo simulations. 

For the photometric analysis we choose individual control sample quasars to have the same redshift (up to $\pm 0.1$) as that of individual primary sample quasars. This ensures that we are not sampling slightly different observed redshift distributions caused by gravitational lensing. Similarly, for the spectroscopic sample we work in narrow redshift bins ($\delta z=0.2$) so that our results are not masked by the smearing of quasar emission lines to continuum-like features  (e.g., York et al. 2006). To test for potential biases we verified that both analyses produce a null mean color excess and a mean spectral ratio of unity, respectively, when comparing different control sub-samples. 

\begin{figure}
\plotone{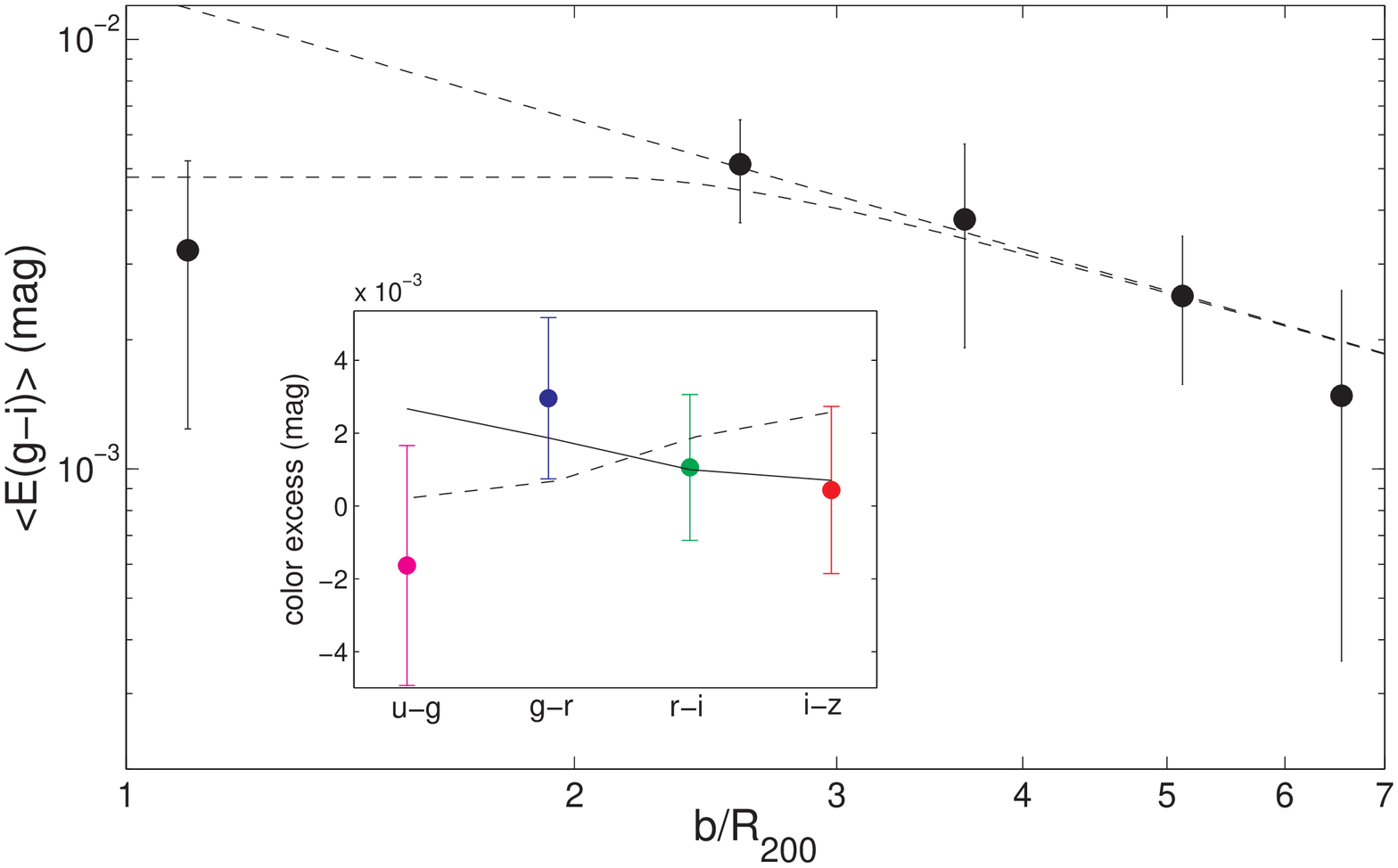}
\caption{$\left < E(g-i) \right >$ as a function of $\tilde{b}$ is consistent with a $\tilde{b}^{-1}$ decline at large radii and a flat core at smaller radii (see overlaid dashed line curves).  The inset shows the results for a particular annulus ($\tilde{b}\sim 5$) for all SDSS colors. $u-g$ is considerably noisier than the other colors (see text). The color excess for $g-r,~r-i,~i-z$ is consistent with known reddening curves (solid line) indicating that foreground emission (dashed line) is negligible  (see also \S 4 and appendix A). }
\label{photo}
\end{figure}

\section{Results}

The color excess is  shown in Figure \ref{photo} as a function of $\tilde{b}$.  Clearly, reddening in $g-i$ is detected with high  significance. The color shifts are consistent with known reddening curves so that $\left < E(g-r) \right > > \left < E(r-z) \right > > \left < E(i-z) \right >$, indicating that foreground emission by the cluster is negligible (see Fig. \ref{photo} and \S4). Our $\left < E(u-g) \right >$ estimate is less secure given the known calibration issues associated with the $u$-band filter\footnote{see e.g., http://www.sdss.org/dr5/algorithms/fluxcal.html}. We find $\left < E(g-i) \right >$ values of  a few$\times 10^{-3}$\,mag.  Inspection of the color-excess distributions (not shown) reveals that the results are not dominated by a few outliers and that most quasars are redder behind clusters. The qualitative agreement with known reddening curves suggests that  $\xi$ is negligible. 

\begin{figure*}
\plotone{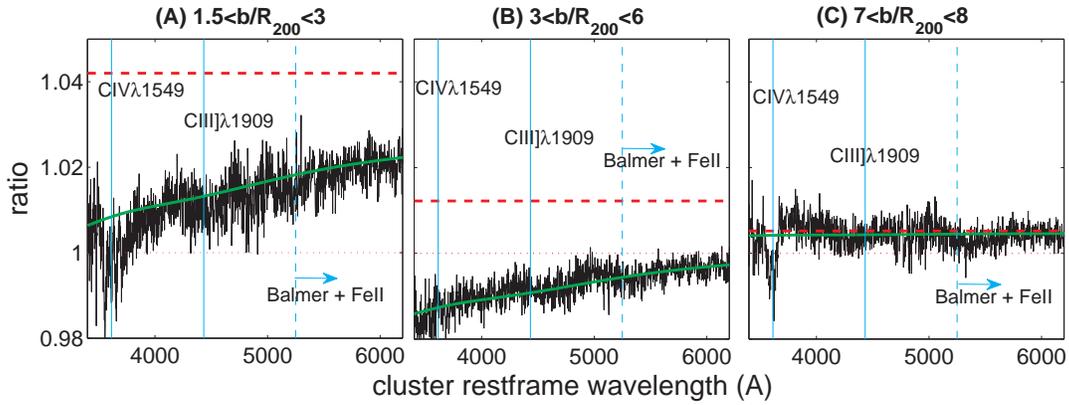}
\caption{The mean spectral ratio for quasars with $1.7<z<1.9$ and at various annuli around the cluster center to  field quasars ($\tilde{b}>7$) at the same redshift bin (the primary samples with $\tilde{b} \in [1.5,3],[3,6]$ consist of $2\times 10^3,~8\times 10^3$ objects, respectively). In all cases the ratio is well fitted by a reddening curve (green line). The dashed red line shows the flux ratio after correcting for the effects of extinction and approaches unity for large $\tilde{b}$, as expected (see text).  Our algorithm does not introduce an artificial signal as seen in the right panel (a low-level signal is observed since the control sample is composed mainly of $\tilde{b} > 8$ quasars). Prominent quasar emission lines are marked as is the wavelength beyond which the Balmer continuum and \ion{Fe}{2}\ line bands are present (see text).}
\label{spec}
\end{figure*}

\begin{figure}
\plotone{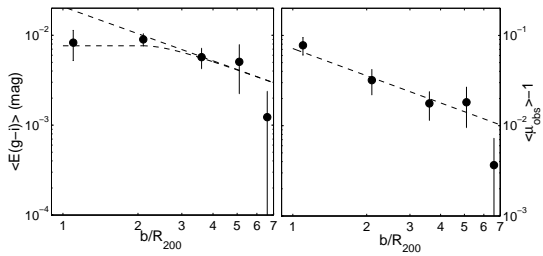}
\caption{Reddening and magnification from the spectroscopic analysis as a function of scale $\tilde{b}$.  {\it Left panel:} Positive $E(g-i)$ (i.e., reddening) is detected with high significance. Here too, reddening rises less sharply toward the center compared to a $\tilde{b}^{-1}$ law and is consistent with a flat core profile (overlaid dashed line curves).  {\it Right panel:} The observed magnification, $\mu_{\rm obs}-1$ as a function of $\tilde{b}$ is of order a several per-cent. No significant deviations can be seen from a $\tilde{b}^{-1}$  magnification profile (dashed line).}
\label{red_mag}
\end{figure}

Motivated by our photometric results we assume that the flux of background quasars is  subject only to extinction and magnification so that the ratio of the observed (flux) spectrum, $F_{\rm obs}$, to the intrinsic one, $F_{\rm int}$, is
\begin{equation}
F_{\rm obs}(\lambda)/F_{\rm int} (\lambda) =\mu e^{-\tau(\lambda)}
\end{equation}
where $\tau(\lambda)$ is given by the extinction curve. In our spectroscopic analysis we consider $1.5<z<2$ quasars which ensures  i) that lensing effects, hence $\xi$, are smaller (see appendix), ii) adequate S/N at each redshift bin (see Fig. 3 in Schneider et al. 2007 showing that $\sim 27$\% of DR5 quasars are in this redshift range), and iii) adequate wavelength coverage excluding the Balmer continuum and emission lines band (which we verified contaminate the signal, especially in lower-$z$ objects;  Yip et al. 2004). A representative case  is shown in Fig. \ref{spec} which is nicely fit by a Cardelli extinction curve (other curves give similar results). Deviations from a smooth reddening curve at the location of the \ion{C}{4}$\lambda 1549$ line are probably a manifestation of the Baldwin (1977) effect (see appendix B). Using $\chi^2$ statistics while avoiding emission line regions and the Balmer and \ion{Fe}{2}\  band, gives us the reddening and the extinction,  allowing us to deduce the extinction-corrected flux ratio i.e., the mean observed gravitational magnification, $\mu_{\rm obs}$ (see Fig. \ref{spec}).  We carried out the analysis for each redshift bin and  averaged our results over all bins to produce Fig. \ref{red_mag}. Typical $\left < E(g-i) \right >$ values are larger than the photometrically deduced values on account of the latter being contaminated by the Balmer/ \ion{Fe}{2} band.  The observed magnification is of order a few per cent.

\section{Conclusions}

We deduce finite extinction  for quasars behind clusters  and our results are consistent with a large covering factor of dust toward these objects. The dust mass enclosed within a sphere of radius $b$ is
\begin{equation}
M_{\rm dust} \sim 10^{9} \left ( \frac{b}{{\rm Mpc}} \right )^2 \frac{\left < E(g-i) \right >}{10^{-2}\,{\rm mag}} ~{\rm M_\odot}.
\end{equation}
(Here we used the Kramers-Kronig relations and grain density of $2.5\,{\rm g~cm^{-3}}$; Kr\"ugel et al. 2003). This  implies a dust-to-gas ratio for the ICM which is $<5\%$ of local inter-stellar medium (ISM; here we used ICM mass of $4\times 10^{13}\,{\rm M_\odot}$ within the central Mpc; Morandi et al. 2007).  If, as suggested by some theoretical models, galaxies eject their ISM in amounts comparable to their own mass during their formation, then our results imply that either the associated dust is in a clumpy form with a small covering factor or else it is rapidly destroyed and the metals are deposited in the ICM. Large scale distributions of dust could arise from a non-virialized gas component at the outskirts of clusters left from past merger events or related to halos of neighboring galaxies and small groups provided their size is of Mpc scales. Singular isothermal sphere (SIS) models, where dust follows dark matter, seem to over-predict the extinction on small  scales (see Figs. \ref{photo} \& \ref{red_mag}) indicating that the centers of clusters may be relatively devoid of ISM-like dust.  

Our results indicate that foreground emission is not important.  This means that the contribution of foreground light that enters a (median) quasar's  point-spread function is $>7$\,mags fainter in the $z$-band (Fig.1 and appendix A) and requires that most sight lines pass $>10$ (de-Vaucouleur) half-light radii ($\sim15$\,kpc; see Fig.  4 in Bernardi et al. 2003) from a median cluster member.  Repeating the entire analysis for quasars whose sight lines pass $>30$\,kpc from confirmed members  yields similar results. We estimate foreground light contamination by resolved and non-resolved cluster members (using a luminosity function with a slope of $-0.8$ extrapolated down to $\sim 0.04L^\star$, and including 25\% of blue galaxies; Hansen et al. 2007) to be at the per cent level, and hence negligible.

The observed magnification was found to be few per cent on Mpc scales. Correcting for luminosity function effects (see appendix B) the intrinsic magnification is of order 10\%, which is in qualitative agreement with SIS model predictions for the magnification of  a $z= 1.7$ quasar behind a rich $z= 0.2$ cluster being $\mu-1\sim 0.2 (b/{\rm Mpc})^{-1}$ (e.g., M\'enard et al. 2007). Given the uncertainties on the dark-matter profile, our approximation for the SDSS selection function and the luminosity function of quasars, we find this agreement to be surprisingly good. It also implies that grey opacity  (by large grains that survived for a Hubble time at the cluster center) is not required.  The above agreement, the fact that the deduced reddening and magnification follow different trends at small $\tilde{b}$ (Figs. \ref{photo} \& \ref{red_mag}), and the good fit to known reddening curves, all suggest that our results are not significantly contaminated by luminosity-dependent intrinsic color differences in the lensed quasar population. The agreement between the photometric and spectroscopic results, which are prone to different systematics, is also reassuring. 

\acknowledgements

We thank Brice M\'{e}nard and Keren Sharon for illuminating discussions and the referee for valuable comments. This research has been supported by NASA through a Chandra Postdoctoral
Fellowship award PF4-50033. DVB is funded through NASA Long Term Space
Astrophysics Grant NNG05GE26G.

\begin{appendix}

\section{Color variations in objects behind  galaxy clusters}

Consider how an intervening cluster  affects a background quasar's flux, $F$  so that the measured flux 
\begin{equation}
F'=F\mu e^{-\tau} + F^{f/g}~\rightarrow~ \tilde{m}_i = m_i+ \frac{2.5}{{\rm ln 10}} \tau_i -2.5{\rm log} ( \mu ) +\delta^F_i
\end{equation}
where $m~(\tilde{m})$ is the (measured) observed magnitude, $\mu$ is the gravitational magnification, $\tau$ the (dust) optical depth through the cluster, and $F^{f/g}$ the  contribution from a foreground source associated with the cluster. Here, $i$ denotes the relevant SDSS filter band and $\delta ^F_i(R)$ the contribution from foreground cluster emission where
\begin{equation}
\delta^F_i \equiv \frac{F^{f/g}}{F} \mu^{-1} e^{\tau} \simeq \frac{F^{f/g}}{F}.
\end{equation}
Defining the mean color of quasars at a given impact parameter $b$ as $\left < i-j \right >_b\equiv \left < \tilde {m}_i(b)-\tilde{m}_j(b) \right >$, then their color excess with respect to field quasars (i.e., at $b \rightarrow \infty$) is
\begin{equation}
 \left < E(i-j) \right >_b \equiv \left < i-j \right >_b-\left < i-j \right >_\infty =  \frac{2.5}{{\rm ln} 10} \left < \tau_i-\tau_j \right >_b + \left < \delta^F_i-\delta^F_j \right >_b + \xi_b.
\label{a3}
\end{equation}
The color excess by foreground emission may be recast in the form
\begin{equation}
\left < \delta^F_i-\delta^F_j \right > = C(m_i,m_j) e^{\Delta m_i}~~~{\rm where}~~~~ C\equiv  2.5 \left [ 1- e^{\left < m^{f/g}_i -m^{f/g}_j \right > - \left <
{m_i-m_j} \right >} \right ]~{\rm and}~~~ \Delta m_i \equiv \left < m_i^{f/g} - m_i \right >.
\end{equation}
Here $m_i^{f/g}$ is due to foreground emission  (see the inset of Fig. \ref{photo} where a $\Delta m_z=7$ model is plotted in dashed line and a typical early-type galaxy spectrum is assumed to characterize the cluster's  spectral energy distribution as a whole; see \S 2). The last term in equation \ref{a3} is: 
$\xi_b  \equiv \left < m_i-m_j \right >_b - \left < m_i-m_j \right >_\infty$ which stands for the {\it "intrinsic"} color difference between quasars behind the cluster and field quasars. This effect can be considerable  {\it if}  (a) gravitational lensing is important {\it and} (b) quasar colors depend on their intrinsic luminosity. The origin of this effect is detailed below.

\section{Mean Properties in Magnitude-limited samples}

Consider a property $Q$ of quasars which depends on their intrinsic luminosity (or magnitude; assume for simplicity a narrow redshift range). In a magnitude limited sample with selection function $\Theta(m-m_l)$ (here we approximate $\Theta$ to be a step function with a limiting magnitude $m_l=19.1$; Schneider et al. 2007; see also M\'enard 2005), the average of $Q$ is 
\begin{equation}
\left < Q  \right > = C_\Psi ^{-1} \int^{\infty}_{-\infty} dm \Theta (m-m_l) Q(m) \Psi(m)~~~;~~~~C_\Psi=\int^{\infty}_{-\infty} dm \Theta (m-m_l) \Psi(m).
\end{equation}
Here $\Psi$ is the luminosity function. The difference between the measured $\left < Q \right >$  when magnification operates (i.e., $m \rightarrow m + \delta m;~\delta m <0$) to when it is not, is to first order
\begin{equation}
\delta \left < Q \right >\simeq C_\Psi ^{-1} \delta m \int^{\infty}_{-\infty} dm \Theta (m-m_l)\left (\frac{\partial Q}{\partial m}\Psi+Q \frac{\partial \Psi}{\partial m} \right ) \ne 0 \rightarrow \frac{\delta \left < Q \right >}{ \left < Q \right >}\simeq (\alpha+\beta)\delta m
\end{equation}
where in the last step we considered a specific example in which  $Q(m)\propto e^{\alpha m}$, and $\Psi(m) \propto e^{\beta m}$. Clearly, the effect on mean quantities can have either sign, depending on the sign of $\alpha+\beta$.  Taking, for example, $\alpha=-0.8$  which characterizes the dependence of quasar emission line luminosity, $L_l$, on the quasar luminosity (via the Baldwin relation; e.g., Dietrich et al. 2002) and $\beta=1$ for the quasar luminosity function, we find that $\delta \left < L_l \right >/\left < L_l \right > \sim -1$\% which is in qualitative agreement with our findings (note the feature at the \ion{C}{4}\,$\lambda1549$ wavelength; Fig. \ref{spec}). Less elegant expressions for geometric means lead to similar results in this case. Given the uncertainties on $\alpha,~\beta$, it would not be surprising to find a different trend  for other lines at other redshifts and luminosity ranges.  Similar considerations apply also for the Balmer continuum and \ion{Fe}{2}\ lines band yet the luminosity dependence is not well known. As for measuring the magnification itself, this corresponds to $\alpha=-1$ which, for $\beta=1$, yields $\delta \left < Q \right >=0$ (i.e., $\mu=1$). Clearly, this is not the case and a more realistic luminosity function to consider deviates from a powerlaw and is of the form $ \Psi(m) \propto \left [10^{-\beta(m-m_\star)}+10^{-\gamma (m-m_\star)} \right ]^{-1}$ with $m_\star=19.1,~\beta=0.98$, and $\gamma=0.15$ (e.g., Myers et al. 2003). Using that, we obtain that the {\it observed} mean magnification is related to the real one by  $\mu-1\sim 3\times (\mu_{\rm obs}-1)$ (see also M\'enard 2005).

\end{appendix}

\end{document}